\documentclass[twocolumn,minimal,showpacs,longbibliography,nofootinbib,pra]{revtex4-1}
\usepackage{graphicx}
\usepackage{textcomp}
\usepackage{amsmath,amssymb}
\usepackage{color}
\usepackage{mathtools}
\usepackage{cleveref}
\usepackage{soul}
\usepackage{xcolor}
\usepackage{subfig}

\crefname{figure}{FIG.}{FIG.}
\Crefname{figure}{FIG.}{FIG.}
\crefname{equation}{eq.}{eqs.}
\Crefname{equation}{Eq.}{Eqs.}

\begin{document}

\title{ Out-of-time-order correlation and detection of phase structure in  Floquet transverse Ising spin system}

\author{Rohit Kumar Shukla}
\email[]{rohitkrshukla.rs.phy17@itbhu.ac.in}
\affiliation{Department of Physics, Indian Institute of Technology (Banaras Hindu University), Varanasi - 221005, India}

\author{Gautam Kamalakar Naik}
\email[]{gautamk.naik.phy15@itbhu.ac.in}
\affiliation{Department of Physics, Indian Institute of Technology (Banaras Hindu University), Varanasi - 221005, India}

\author{Sunil Kumar Mishra}
\email[]{sunilkm.app@iitbhu.ac.in}
\affiliation{Department of Physics, Indian Institute of Technology (Banaras Hindu University), Varanasi - 221005, India}
\begin{abstract}
We study the out-of-time-order correlation (OTOC) of the Floquet transverse Ising model and use it to verify the phase diagram of the system.
First, we present the exact analytical solution of the transverse magnetization OTOC using the Jorden-Wigner transformation.  We calculate the speed of correlation propagation and analyze the behavior of the revival time with the separation between the observables. In order to get the phase structure of the Floquet transverse Ising system, we use the longitudinal magnetization OTOC as it is known to serve as an order parameter of the system. We show the phase structure numerically in the transverse Ising Floquet system by using the long time average of the longitudinal magnetization OTOC. In both the open and the closed chain systems, we find distinct phases out of which two are paramagnetic (0-paramagnetic and $\pi$-paramagnetic), and two are ferromagnetic (0-ferromagnetic and $\pi$-ferromagnetic) as defined in the literature.
\end{abstract}


\maketitle
\section{Introduction}
In the last two decades, out-of-time-order correlation (OTOC) has gained a lot of attention among the researchers of various fields. One field of interest is the butterfly effects in quantum chaotic systems
\cite{Ray2018, Hosur2016, gu2016,ling2017}.  Other directions are quantum information scrambling \cite{Bohrdt2017,Yao2017,Swingle2016,Swingle2017,Schleier2017,Pappalardi2018,Klug2018,Khemani2018,Hosur2016,Alavirad2018} and many-body localization  \cite{Maldacena2016}. The nontrivial OTOC as a holographic tool  has been instrumental in determining the interplay of scrambling and entanglement  \cite{Shenker2014,Roberts2015}. Many experiments have been done to measure OTOCs in various systems, {\it e.g.}, trapped-ion quantum magnets \cite{Garttner2017} and Nuclear Magnetic resonance quantum simulator \cite{Li2017}.  
 \par  
In addition to the above fields of interest, the OTOCs are useful in determining phases of the quantum critical systems \cite{Shen2017,Sun2020,Heyl2018}. The phase structures of quantum critical systems have been studied extensively in the last few decades \cite{Shen2017,Heyl2013,Pollmann2010,von2016phase,Thakurathi2013,Turner2011,Keyserlingk2016a,Feng2007,Bastidas2012,Jiang2011,Sun2020,Fidkowski2011,Khemani2016,Thakurathi2014}. One of the simplest models to display and analyze the quantum phase transition is one dimensional transverse Ising model which is given by Hamiltonian $H=J\sum_{i}\sigma_i^x\sigma_{i+1}^x+h\sum_i\sigma_i^z$. This system undergoes a phase transition at $J=h$ from the ferromagnetic state ($J>h$) to the paramagnetic phase ($J<h$) \cite{chakrabarti2008,Heyl2018,su2006,sun2009}. Such phase transitions in time-independent equilibrium systems have been well studied over the years. In the last few years, the OTOC has emerged as a tool to detect equilibrium and dynamical quantum phase transitions in the transverse field Ising (TFI) model and the Lipkin-Meshkov-Glick model (LMG)  \cite{Heyl2018}.
\par
\begin{figure}[t!]
\centering
\includegraphics[width=\linewidth, height=.9\linewidth]{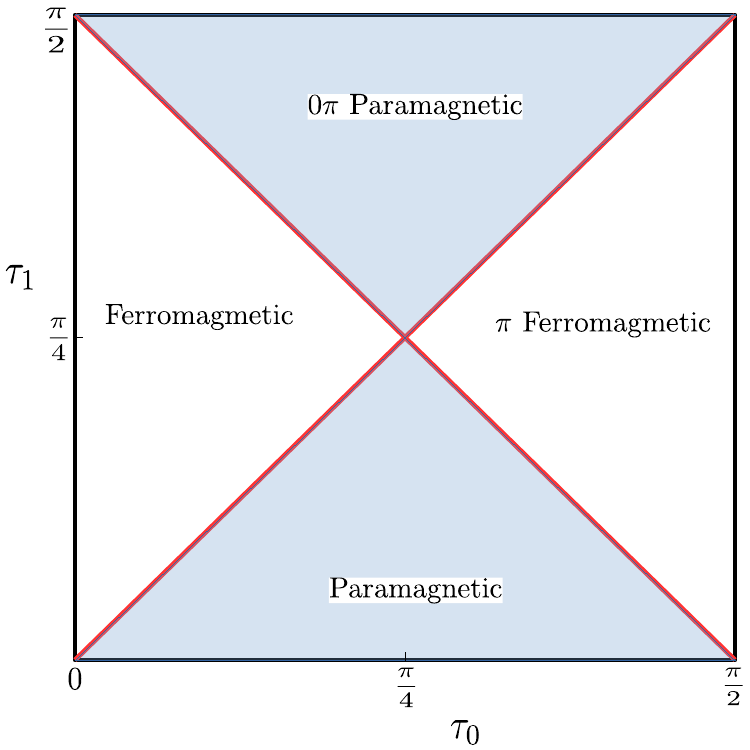}
\caption{Phase structure of the Floquet system with Floquet map given by \cref{U_f} \cite{Khemani2016,von2016phase,Keyserlingk2016a}. There are four distinct phases in the $\tau_0$-$\tau_1$ parameter space. Two of these phases, the $\pi$ ferromagnetic and the $0\pi$ paramagnetic, and phases which are unique to Floquet systems.}
    \label{four_phase}
\end{figure}

  
It has been shown that the OTOC of the ground states and quenched states can diagnose the quantum phase transitions and dynamical phase transition, respectively \cite{Heyl2018}. The ferromagnetic ($J>h$)and paramagnetic ($J<h$) phases of the transverse Ising model can be characterized by nonzero and zero long time averaged OTOC, respectively\cite{Heyl2018}. Periodically driven quantum systems, known as the Floquet systems which have properties of the duality between time and space \cite{akila2016particle} and time-reflection symmetry \cite{iadecola2018floquet}, on the other hand, poses a different problem: one would expect generic Floquet systems to heat to infinite temperatures. However, specific cases of nonergodic phases with localization have been observed in Floquet systems \cite{Parameswaran2017,zhang2016}. In these systems, multiple nonergodic phases with differing forms of dynamics and ordering have been observed \cite{Khemani2016}. These multiple phases are characterized by broken symmetries and topological order. In the case of transverse Ising Floquet systems, Majorana modes are produced at the ends of the  chain \cite{Thakurathi2013}. The zero-energy Majorana mode corresponds to the long range ferromagnetic order while the nonzero-energy Majorana mode corresponds to the paramagnetic phase \cite{Thakurathi2013, von2016phase}. The sharp phase boundaries of the Floquet Ising system are explained by symmetry protected Ising order  \cite{ Khemani2016, Keyserlingk2016a}.  For binary Floquet drive, two paramagnetic and two ferromagnetic phases can be seen in the phase diagram.  The two paramagnetic/ferromagnetic phases are distinguished by the combined eigenvalues at the edges of the Floquet drives and the parity operators. On the basis of the combined eigenvalues, the paramagnetic region is divided into two parts: $0$ and $0\pi$-paramagnetic, and ferromagnetic region is also divided into two parts: 0 and $\pi$-ferromagnetic [\cref{four_phase}].  In the ferromagnetic region, all of the eigenstates have long range Ising symmetry broken order. However, in the paramagnetic phase, all of the eigenstates have long range symmetric order. 
 \par
First, we will consider transverse magnetization out-of-time-order correlation (TMOTOC) and calculate the exact solution using the Jordan Wigner transformation by mapping the spin operators onto the fermionic annihilation and creation operators. Next, we will consider the longitudinal magnetization out-of-time-order correlation (LMOTOC) and explore the various phases in the Fouquet Ising spin system. 
\par
In this letter we will start  discussing the model of the Floquet system. Subsequently, we will define the longitudinal and transverse magnetization  OTOCs.  We will introduce the time average of the  LMOTOC for the detection of phase structures and discuss the various phases of the Floquet Ising system using a long time averaged LMOTOC. Later, we conclude the results.
\section{Model}
\label{model}
We consider an integrable Floquet transverse Ising system with binary Floquet drives. The Floquet map corresponding to this system is (taking $\hbar=1$)  
\begin{equation}
\label{U_f}
U=e^{-iH_{xx}\tau_{1}}e^{-iH_{z}\tau_{0}},
\end{equation} 
where $H_{xx}$ is the nearest neighbor Ising interaction given by $H_{xx}=\sum_{l=1}^{N-1}\sigma_l^x\sigma_{l+1}^x $ for open chain system and $H_{xx}=\sum_{l=1}^{N}\sigma_l^x\sigma_{l+1}^x $ for closed chain system with $\sigma_{N+1}^x=\sigma_{1}^x$.
$H_z=\sum_{l=1}^{N}\sigma_l^z$ is the transverse field in z-direction.  

\section{ Out-of-time-order Correlation}
  \label{avgOTOC}
The out of time order correlation (OTOC) is, in general, defined as: $F(t)=\langle W(t) V W(t) V \rangle $,
where $V$ and $W$ are two local Hermitian operators and $W(t)$ is the Heisenberg evolution of the operator $W$ by time $t$.
In the manuscript we consider two different OTOCs defined as:
\begin{enumerate}
    \item[i)] {\it Transverse magnetization OTOC (TMOTOC) }: 
Here we consider two local spin operators $W$ and $V$ in the direction perpendicular to  the Ising axis (x-axis). In our generic treatment we set the operators $W=\sigma_l^z$ and $V=\sigma_m^z$ at different sites $l$ and $m$.   The TMOTOC in our protocol is given as: 
\begin{eqnarray}
\label{F_z}
F_z^{l,m}(n) &=&  \langle \phi_0|\sigma_l^z(n)\sigma_m^z\sigma_l^z(n)\sigma_m^z|\phi_0\rangle, 
\end{eqnarray}
with the initial state as $|\phi_0\rangle=| \uparrow  \uparrow  \uparrow \cdots  \uparrow \rangle,$ 
where $ \left| \uparrow \right\rangle$ is the eigenstate of $\sigma^z$ with eigenvalue $+1$.
    \item[ii)] {\it Longitudinal magnetization OTOC (LMOTOC) }:
    In this case two local spin operators are chosen along the Ising axis i.e. $W=\sigma_l^x$ and $V=\sigma_m^x$. The LMOTOC is given as:
\begin{equation}
    \label{F_x}
F_x^{l,m}(n) =  \langle
 \psi_0|\sigma_l^x(n)\sigma_m^x\sigma_l^x(n)\sigma_m^x|\psi_0\rangle.
\end{equation}
Here $|\psi_{0} \rangle=|\rightarrow \rightarrow \rightarrow \cdots \rightarrow \rangle$ is the initial state 
with, $ \left| \rightarrow \right\rangle$ is the eigenstate of $\sigma^x$ with eigenvalue $+1$. 
\end{enumerate}
In what follows $l$ and $m$ can take any value between $1$ to $N$ (even) in closed chain system. For the open chain case we will consider the special case with $l=m=\frac{N}{2}$. 
The time evolution of the spin operator at  the position $l$ after $n$ kicks is defined as $\sigma_l^{z/x}(n)=U^{\dagger n} \sigma_l^{z/x} U^n$.
The case $l=m$ will be treated as a special case.
Considering  $t_0=2 \tau_0$, $t_1=4 \tau_1$ and $\sigma_l^x=2 S_l^x$ and periodic boundary condition in the  unitary operator defined in \cref{U_f}, we get Floquet map as:
\begin{eqnarray}
 U &=&\exp\Big[-i t_1 \sum_{l=1}^{N}S_l^x S_{l+1}^x \Big] \exp\Big[-i t_0 \sum_{l=1}^{N} S_l^z\Big],
\end{eqnarray}
 
  We calculate the analytical expression for the TMOTOC using the Jorden-Wigner transformation (for detailed calculation refer to the Supplementary Material S-I [Supplementarymaterial.pdf] ):
 \begin{eqnarray}
\label{OTOCz_gene}
 F_z^{l,m}(n) &=& 1- \Big(\frac{2}{ N}\Big)^3 \sum_{p,q,r} \Big[ e^{i(p-q)(m-l)}|\Psi_r(n)|^2 \Phi_p^{*}(n)  \nonumber \\
 &\times& \Phi_q(n) + e^{i(q-r)(m-l)} \Psi_{q}(n)\Psi_r^*(n) |\Phi_p(n)|^2 \nonumber \\ 
 &-& e^{i(-r-q)(m-l)}  \Psi_r(n)^{*} \Phi_p^{*}(n)  \Phi_q(n)\Psi_{-p}(n)   \nonumber \\
&-& e^{i(p+q)(m-l)} \Psi_{q}(n)  \Psi_{r}^{*}(n)  \Phi_{p}^{*}(n) \Phi_{-r}(n)  \Big].
\end{eqnarray}
Now, we take a special case in which both the local operators are at the same position i.e. $l=m$ and $V=W=\sigma_l^z$. The expression of  TMOTOC simplifies to  
 \begin{eqnarray}
   \label{OTOCz}
 F_z^{l,l}(n) &=& 1- \Big(\frac{2}{ N}\Big)^3  \sum_{p,q,r}\Big[|\Psi_r(n)|^2 \Phi_p(n)^* \Phi_q(n)  
 \nonumber \\
&-& \Psi_{-p}(n)\Psi_r(n)^* \Phi_p(n)^* \Phi_q(n)-\Psi_q(n)  \Psi_r(n)^*   \nonumber  \\
 &\times &\Phi_{p}(n)^* \Phi_{-r}(n) + \Psi_{q}(n)  \Psi_r(n)^* |\Phi_p(n)|^2 \Big],  
 \end{eqnarray}
where the expansion coefficients  $\Phi_q(n)$ and  $\Psi_q(n)$ are defined as 
\begin{equation}
\label{phi}
\Phi_q(n)=|\alpha_{+}(q)|^2 e^{-i n \gamma_q}+|\alpha_{-}(q)|^2 e^{i n \gamma_q},
\end{equation}
\begin{equation}
\label{psi}
\Psi_q(n)=\alpha_{+}(q) \beta_{+}(q)e^{-i n \gamma_q}+\alpha_{-}(q)\beta_{-}(q) e^{i n \gamma_q}.
\end{equation}
The phase angle $\gamma_q$ and the coefficients $\alpha_{\pm}(q)$ and $\beta_{\pm}(q)$ are given by
\begin{equation}
\label{gamma}
\cos(\gamma_q)=\cos(t_0)\cos(t_1)-\cos(q)\sin(t_0)\sin \Big(\frac{t_1}{2} \Big),
\end{equation}
and
\begin{equation}
\label{apmq}
\alpha_{\pm}(q)^{-1}=\sqrt{1+\Big(\frac{\cos(\frac{t_1}{2})-cos(\gamma_q \pm t_0)}{\sin(q) \sin( t_0)\sin(\frac{t_1}{2})}\Big)^2},
\end{equation}

\begin{eqnarray}
\label{bpmq}
\beta_{\pm}(q)&=& \frac{\mp\sin(\gamma_q)-\cos( t_0) \cos(q) \sin(\frac{t_1}{2})-\sin( t_0) \cos(\frac{t_1}{2})}{\sin(q)\sin(\frac{t_1}{2})}  \nonumber \\
&\times & \alpha_{\pm}(q)e^{-it_0}.
\end{eqnarray}
 It should be noted that in all the calculation we set the lattice constant (say `$a$') to be equal to $1$. The system size of $N$ spins means that the length of the chain is equal to $N$ (in the units of $a$). In the momentum space, the allowed values of $p$, $q$ and $r$ are from $\frac{-(N-1)\pi}{N}$ to $\frac{(N-1)\pi}{N} $ differing by $\frac{2 \pi}{N} $ for even number of $N_F$ ($N_F=c_l^\dagger c_l$, number of fermions), is consistent with lattice constant ($a=1$).
 
\begin{figure}[h!]
     \centering
        \includegraphics[width=\linewidth, height=.6\linewidth]{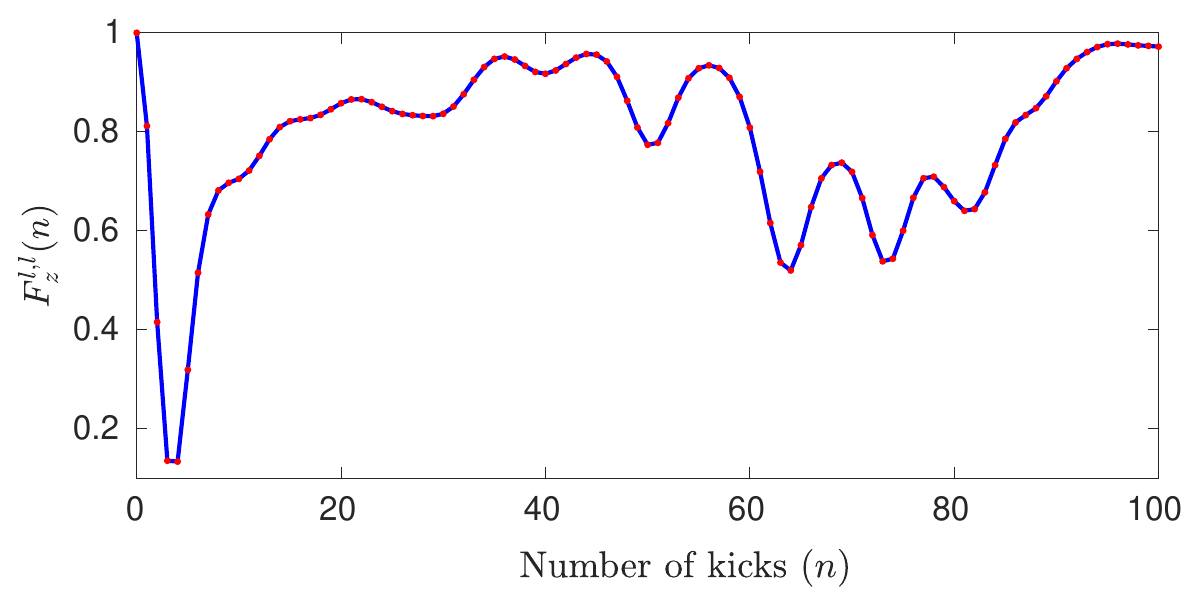}
       \caption{$F^{l,l}_z(n)$ for closed chain transverse Ising Floquet system of system size $N=12$ by using the numerical calculations (solid line) and analytical expression of eq.\ref{OTOCz} (point). Here we take $\tau_0=\tau_1=\epsilon$, where $\epsilon=\frac{\pi}{28}$.}
    \label{comp_otoc_n12}
\end{figure}
\begin{figure}[hbt!]
\centering
  \includegraphics[width=\linewidth, height=.6\linewidth]{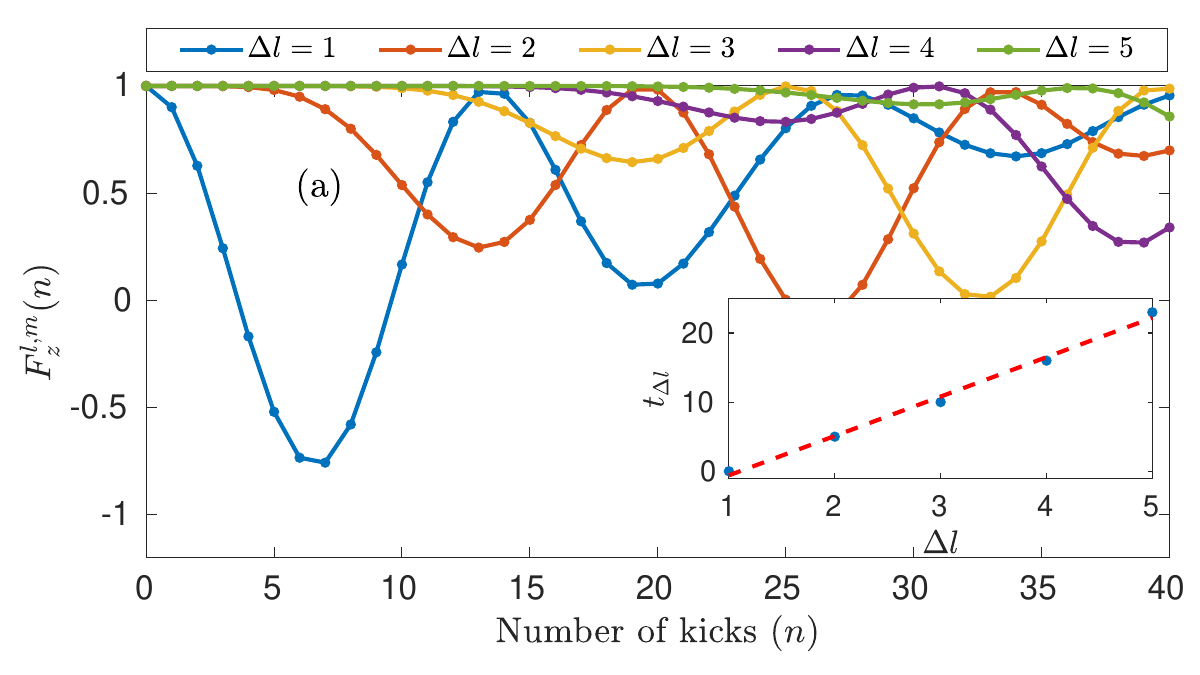}
   \includegraphics[width=\linewidth, height=.60\linewidth]{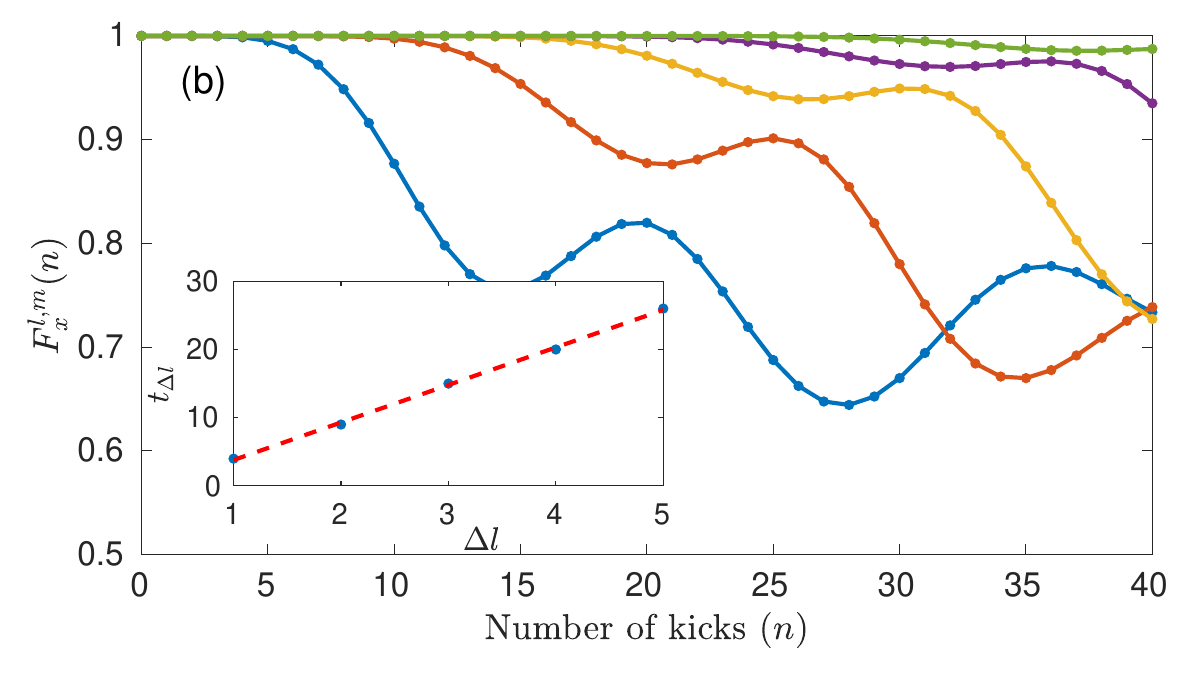}
  
\caption{Behaviour of (a) $F_z^{l,m}(n)$ and (b)$F_x^{l,m}(n)$  with number of kicks for different value of $\Delta l$. Here, the parameters are: $\tau_0=\frac{\epsilon}{2}$,  $\tau_1=\epsilon$,and  $N=12$.  In the both the figures (a) and (b), the inset shows the behavior of time of departure from unity ($t_{\Delta l}$) as a function of separation between the observables ($\Delta l$).}
\label{otocz_11_butterfly}
\end{figure}
 \begin{figure}[hbt!]
     \centering
        \includegraphics[width=\linewidth, height=.60\linewidth]{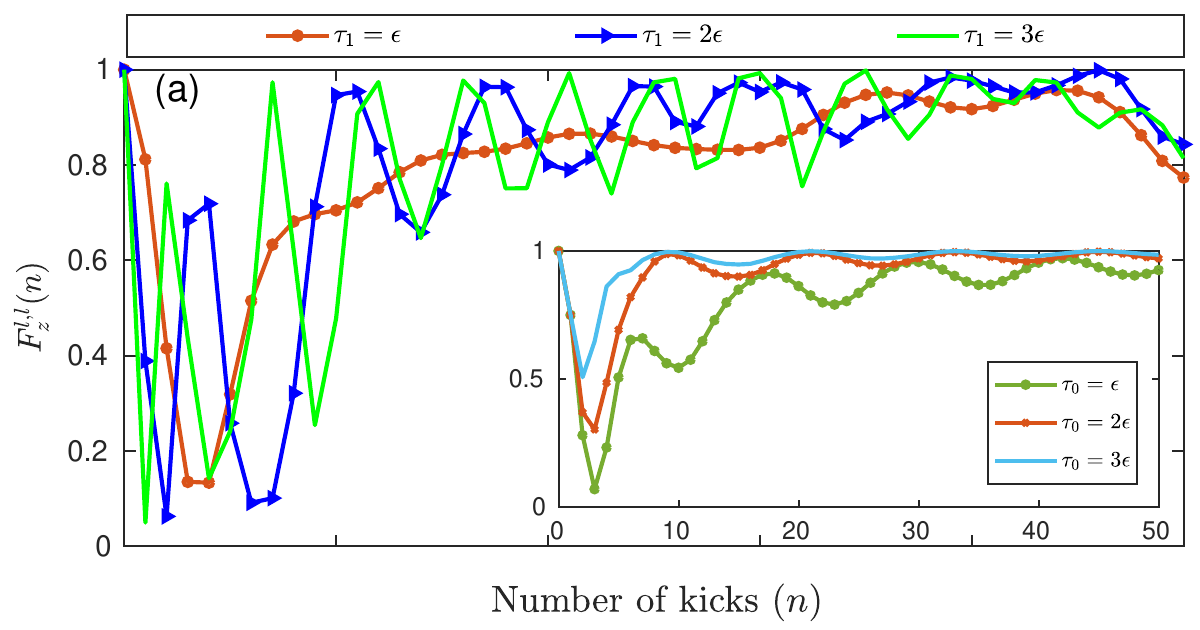}
        \includegraphics[width=\linewidth, height=.60\linewidth]{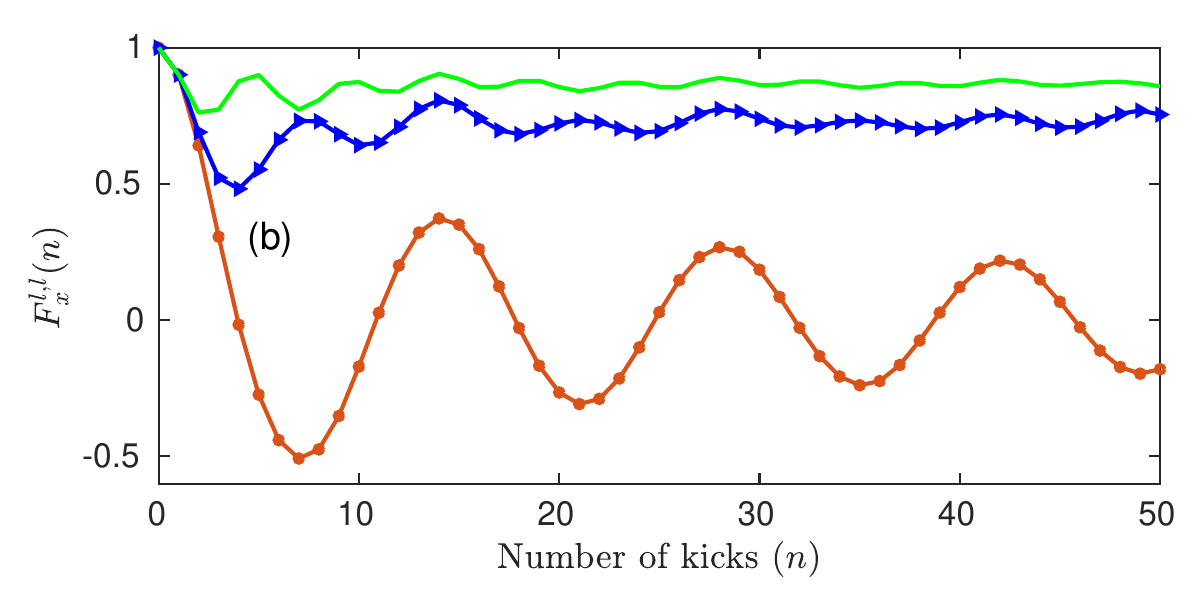}
       \caption{(a) Variation of the real part of (a) $F_x^{l,l}(n)$ and (b) $F_x^{l,l}(n)$ with number of Floquet periods for a fixed $\tau_0=\epsilon$ and $\tau_1 =\epsilon,$ $ 2 \epsilon$ and $ 3\epsilon$ in closed  chain Floquet systems with system size N=12.
Inset of the figure is the variation of $F_z^{l,l}(n)$ with the number of Floquet periods for a fixed $\tau_1=\frac{\pi}{24}$ and $\tau_0 =\epsilon,$ $ 2 \epsilon$ and $ 3\epsilon$ in closed chain Floquet systems with system size N=50.}
    \label{otocz_otocx_n12_tau0_pi28_tau1_pi6}
\end{figure}
The values of $F_z^{l,l}(n)$ obtained  by the analytical expression in eq. \ref{OTOCz} exactly match with those obtained by numerical exact diagonalization as shown in \cref{comp_otoc_n12}.
\par
 Next, we use  eq.\ref{OTOCz_gene} for $l\ne m$ to calculate the speed for correlation propagation. At $t=0$, both the operators $W(t=0)=\sigma_l^z$ and $V=\sigma_m^z$, commute with each other which implies that $F_z^{l,m}(n)$ will be unity.  As time changes, evolution of $\sigma_l^z$ takes place by the Floquet operator, they no longer commute. Therefore, $F_z^{l,m}(n)$ starts to drop from unity which  provides us the speed of correlation propagation ($v_{cp}$). The general approach to calculate $v_{cp}$ in this manuscript is as follows: First, we fix $m=\frac{N}{2}$ and change $l$ from $\frac{N}{2}+1$ to  $\frac{N}{2}+5$. By using {\ref{otocz_11_butterfly}}(a), we determine the characteristic time $t_{\Delta l}$ in which  $F_z^{l,m}(n)$ starts departing from unity and plot it as a function of the separation between the observables ($\Delta l$) [inset of [{\ref{otocz_11_butterfly}}(a)].  In the inset of {\ref{otocz_11_butterfly}}(a), dots are the points corresponding to the given $\Delta l$ and dashed line is the best fit line.  Reciprocal of the slope of this straight line is the  speed of the correlation propagation $v_{cp}$. For comparison, we have  shown similar results for LMOTOC  in {\ref{otocz_11_butterfly}}(b). We find that the speed of the commutator growth  of the $F_z^{l,m}(n)$ ($v_{cp}=0.175$) is nearly equal to that of the $F_x^{l,m}(n)$ ($v_{cp}=0.181$). This means that $v_{cp}$ is independent of the choice of the  observables. By comparing {\ref{otocz_11_butterfly}}(a) and (b), we observe that closer the operators $V$ and $W$ are, smaller the characteristic time $t_{\Delta l}$ is.
\par
 Now we move to another interesting quantity, the revival time of $F_z^{l,m}(n)$ and $F_x^{l,m}(n)$ which is defined as the time in which OTOCs return back to their initial value. We can see from \cref{otocz_11_butterfly}(a,b) that the early time behavior of both $F_z^{l,m}(n)$ and $F_x^{l,m}(n)$  looks very similar. For instance, both $F_z^{l,m}(n)$ and $F_x^{l,m}(n)$ start deviating from unity after a certain time. However, the long time behaviors of $F_z^{l,m}(n)$ and $F_x^{l,m}(n)$ differ widely. After decreasing from unity  to a minimum, $F_z^{l,m}(n)$ revives and recover to its initial value {\it i.e.,} unity, while $F_x^{l,m}(n)$ oscillates about a finite value and  never reaches to unity. Revival time depends on the distance between local operators. The larger the separation between the operators $V$ and $W$ is, the more the revival time is.  This can be seen from {\cref{otocz_11_butterfly}}(a,b).
 \par
The advantage of having an easily computable formula such as eq. (\ref{OTOCz}) is that we can study the TMOTOC  as a function of Floquet periods $\tau_0$ and $\tau_1$ and see the behavior at any number of kicks. The analytical expression is of $O(L^3)$ which has a significant advantage over exact diagonalization calculations of  O($2^L$) [inset of {\cref{otocz_otocx_n12_tau0_pi28_tau1_pi6}}(a)].
\par
The LMOTOCs have been shown to be useful in detecting the phase transitions between the paramagnetic and ferromagnetic phases in spin systems \cite{Heyl2018}. 
However, the same cannot be said about TMOTOC by using the same concept.  A comparison of the behavior of the two quantities with time is shown in \cref{otocz_otocx_n12_tau0_pi28_tau1_pi6}. We see from \cref{otocz_otocx_n12_tau0_pi28_tau1_pi6}(a) that  TMOTOCs always oscillate about a  positive value for all the pairs of $\tau_0$ and $\tau_1$, signalling that the long time average of  TMOTOC is always a positive quantity. However, in the case of  LMOTOCs, as shown in \cref{otocz_otocx_n12_tau0_pi28_tau1_pi6}(b), we find that the long time average value can be either zero or a positive quantity. In order to detect the phase structure of the system, we require the order parameter characterizing the distinct phases to show a sharp contrast between the phases. We see that the  LMOTOCs qualify the criterion to be used as an order parameter, but TMOTOCs fail to do so.

Upon performing a quantum quench from a polarized state, we will use the saturation value of the  LMOTOC as the order parameter to distinguish between the two phases. It is calculated by numerical methods because a compact analytical expression is not achievable using Jordan-Wigner transformation.
A possible approach and inability to get a compact analytical solution for LMOTOC is given in the Supplementary Material S-II (Supplementarymaterial.pdf).
\label{phase_structure}
\begin{figure*}[hbt!]
     \centering
        \includegraphics[width=\linewidth, height=.40\linewidth]{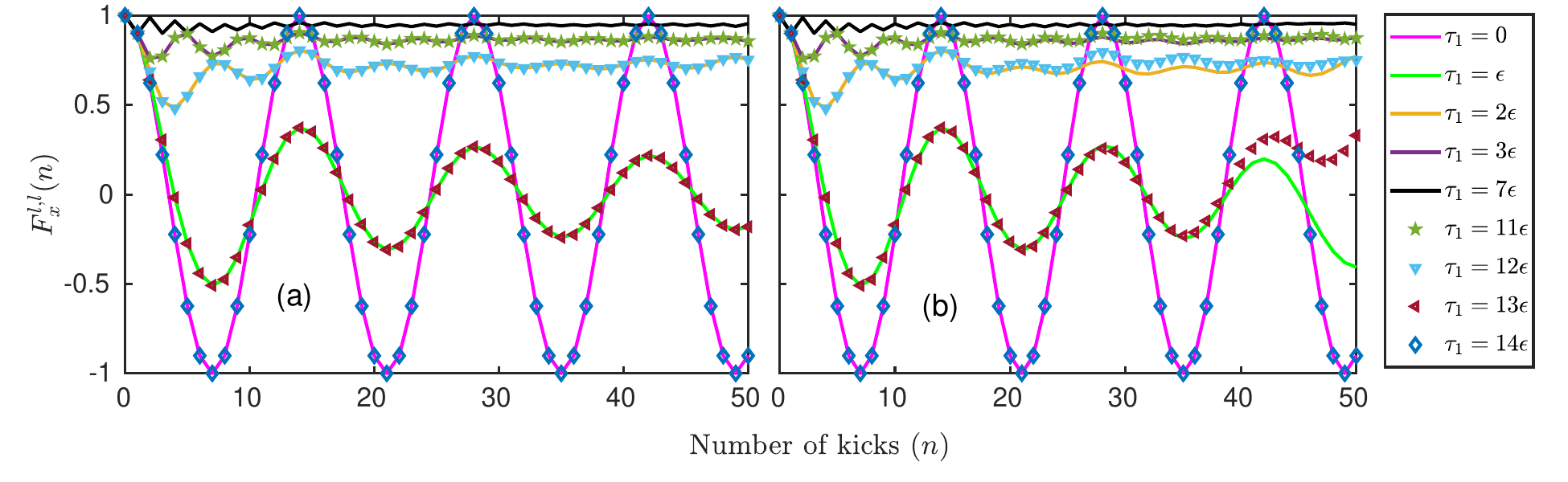}
       \caption{Variation of the real part of $F_x^{l,l}(n)$) with number of Floquet periods for a fixed $\tau_0=\epsilon=\frac{\pi}{28}$ and multiple values of $\tau_1$ in (a) closed and (b) open chain Floquet systems with system size N=12. The initial state is direct product of eigenstate of $\sigma^{x}$ with eigenvalue $+1$. At $\tau_1=0$, $F_x^{l,l}(n)$ shows periodic oscillations about zero. As the value of $\tau_1$ is increased, the $F_x^{l,l}(n)$ oscillates about greater non-zero values with lower amplitudes of oscillation and at $\tau_1=\frac{\pi}{4}$, it saturates to the value $1$. As increase constant value of  $\tau_1$, $F_x^{l,l}(n)$ has same and different value as $ \frac{\pi}{2}-\tau_1$ in closed  and open chain respectively.}
    \label{otocx_tau1_n12_p1_p0_new}
\end{figure*}
We study the phase structure of the system [\cref{U_f}] by calculating the  LMOTOC[\cref{F_x}]. If the  LMOTOC  saturates to a particular value after a long period of time, this value can be determined by taking the long time average of the LMOTOC. We define the long time average of the  LMOTOC, $\overline{F}_x^{l,l}(T)$ upto $T$ Floquet periods by: 
\begin{equation}
\overline{F}_x^{l,l}=\frac{1}{T} \sum_{n=1}^T F_x^{l,l}(n).
\end{equation}
 The long time average of the LMOTOC has a direct link with the spectral properties of the system in consideration \cite{DeMelloKoch2019}. The averaged  LMOTOC links with the spectral form factor \cite{Dyer2017}, a well-known quantity in random matrix theory, which is a quantifier for discreteness in the spectrum.  
\section{Phase Structure}
\begin{figure*}[hbt!]
    \centering
\includegraphics[width=\linewidth, height=.40\linewidth]{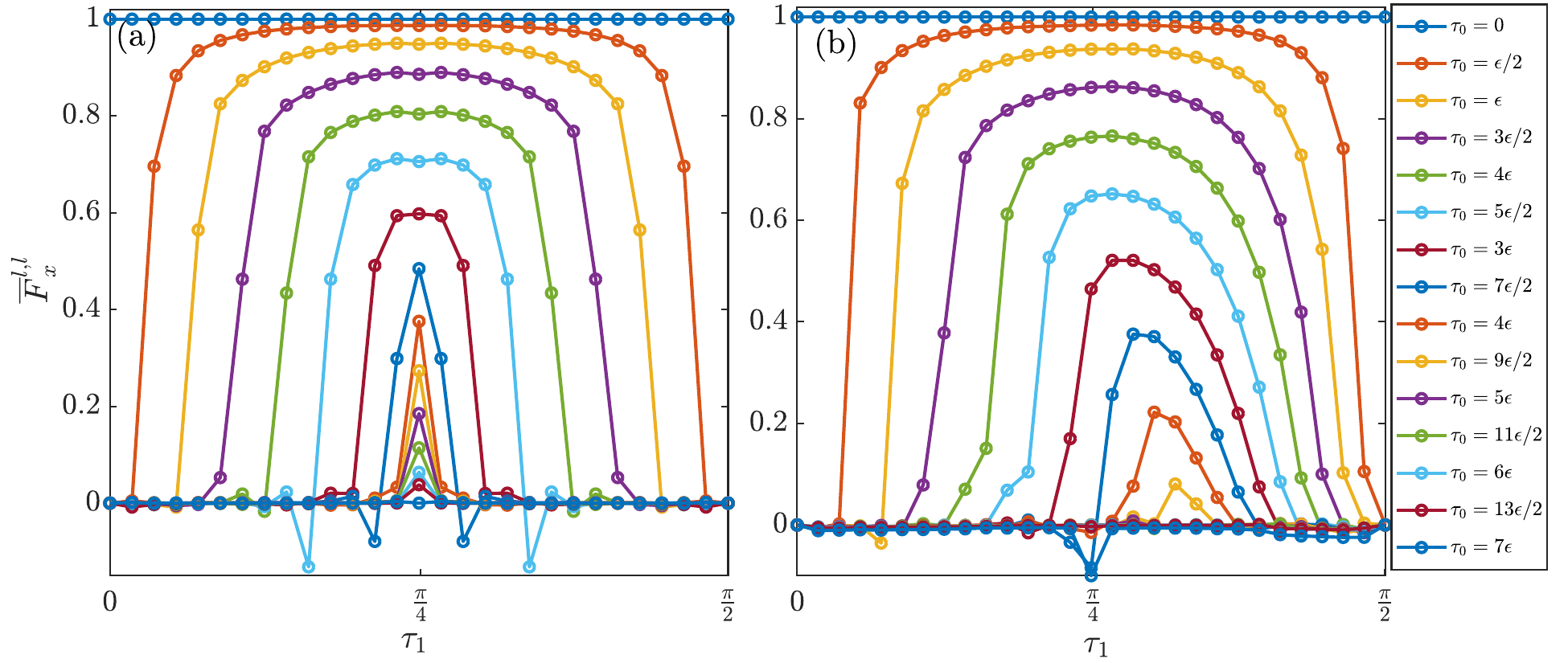}
\caption{ Plot of $\overline{F}_x^{l,l}(T)$ with $\tau_1$ for values of $\tau_{0}$ varying from $0$ to $\frac{\pi}{4}$  in intervals of $\epsilon$ in the (a) closed and (b) open chain Floquet systems of system size $N=10$ and $T=10^4$. The variation of $\overline{F}_x^{l,l}(T)$ with $\tau_1$ for $\pi/4<\tau_0<\pi/2$ is the same as that for $\frac{\pi}{2}-\tau_{0}$. This plot can be used to find the regions in the $\tau_0$ and $\tau_1$ parameter space that have $\overline{F}_x^{l,l}(T)>0$ and $\overline{F}_x^{l,l}(T)=0$ (\cref{finite_phase}).}
    \label{otoc2_n10_tau0}
\end{figure*}
\begin{figure*}[hbt!]
    \centering
    \includegraphics[width=\linewidth, height=.45\linewidth]{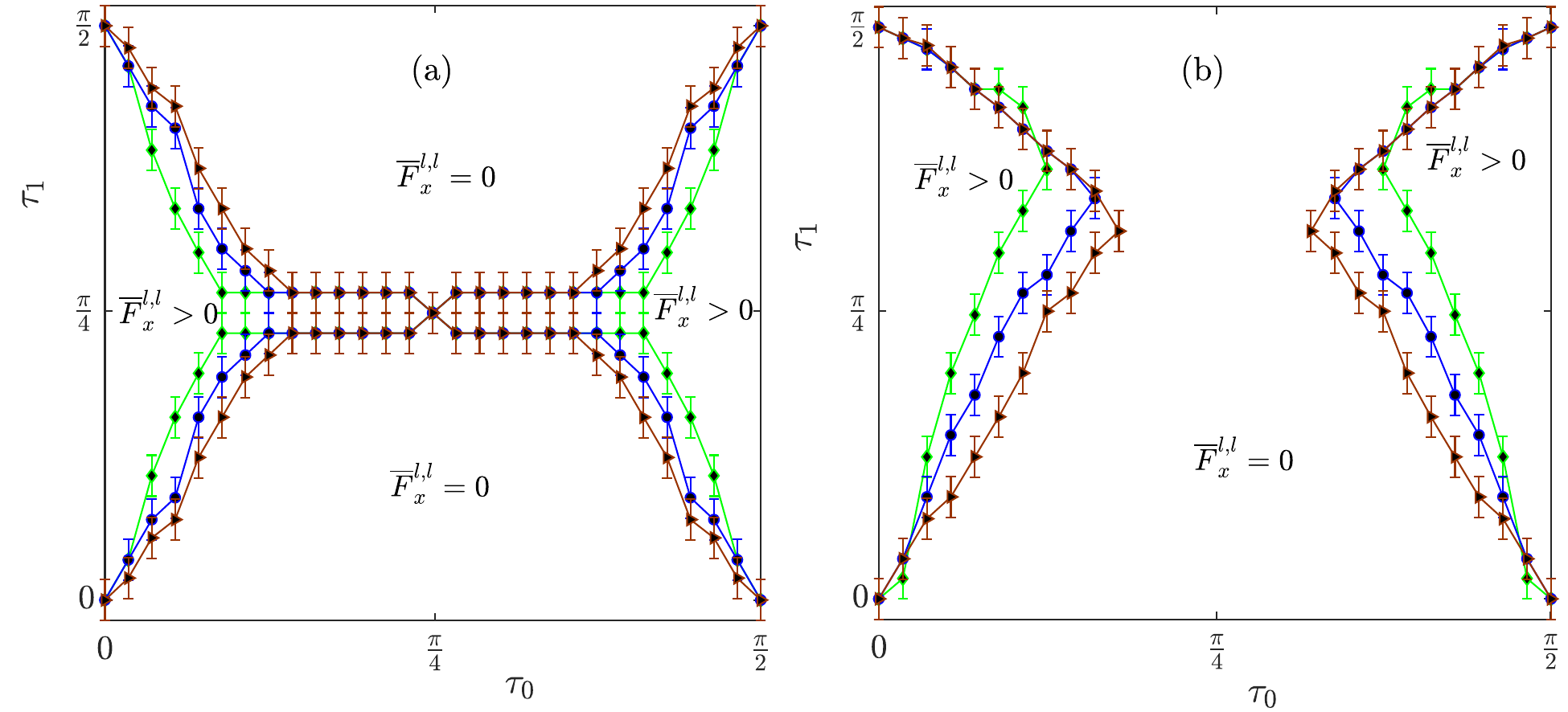}
    \caption{Regions with $\overline{F}_x^{l,l}(T)>0$ and $\overline{F}_x^{l,l}(T)=0$ in the $\tau_0$ and $\tau_1$ parameter space with $T=10^4$, for the (a)closed and (b) open chain Floquet systems of system size N=6(Green), 8(Blue) and 10(Brown). As increasing system size, critical lines of  both (a) and (b) tend towards diagonal critical lines. Hence, this suggests that the phase structure of the system would contain two ferromagnetic regions (where $\overline{F}_x^{l,l}>0$) and two paramagnetic regions (where $\overline{F}_x^{l,l}=0$).}
    \label{finite_phase}
\end{figure*}
The phase structure of the Floquet system given by \cref{U_f} is known to have four distinct phases in the two-dimensional parameter space of $\tau_0$ and $\tau_1$. The phase diagram is shown in \cref{four_phase} \cite{Khemani2016,von2016phase,Keyserlingk2016a}. Paramagnetic and ferromagnetic phases show behavior similar to their undriven counterparts. The other two of the phases observed, the $\pi$-ferromagnetic and $0\pi$-paramagnetic are unique to Floquet systems and are not observed in undriven non-Floquet systems \cite{lin2018out}. The phase transitions between these phases in the $\tau_0$ and $\tau_1$ parameter space can be detected by calculating LMOTOCs. The  LMOTOC has been shown to saturate to non-zero values in the ferromagnetic region and to zero in the paramagnetic region, at long times in the undriven systems \cite{Heyl2018,lin2018out}. Hence, the long time averaged LMOTOC serves as a good order parameter for paramagnetic and ferromagnetic regions in the undriven systems. In driven Floquet systems, LMOTOCs do not saturate to non-zero and zero values for all values of $\tau_0$ and $\tau_1$; we see a continuing oscillating behaviour about a non-zero or zero mean value (\cref{otocx_tau1_n12_p1_p0_new}). The time-averaged LMOTOC ($\overline{F}_x^{l,l}(n)$) is seen to saturate at long times in the thermodynamic limit to non-zero values in the ferromagnetic and $\pi$ ferromagnetic regions, and to zero in the paramagnetic and $0\pi$ paramagnetic regions of the phase space. \Cref{otoc2_n10_tau0} shows the variation of the long time-average of the real part of the LMOTOC  with $\tau_1$, for different values of $\tau_0$ in the closed and open boundary conditions for a system size $N=10$.\\ The critical points, where the phase transition occurs, are identified along the constant $\tau_0$ line at the points where LMOTOC  goes from zero to non-zero. These critical points, when mapped in the $\tau_0$ and $\tau_1$ parameter space for $ N=6, 8$ and $10$ give plots as shown in \cref{finite_phase}.\\
There exists a symmetry along $\tau_1=\frac{\pi}{4}$ in the closed chain case because the behavior of LMOTOC is same for Floquet period $\tau_1$ and $\frac{\pi}{2}-\tau_1$ (e.g., $\epsilon$ and $13\epsilon$, $2\epsilon$ and $12\epsilon$, $3\epsilon$ and $11\epsilon$  are same in   \Cref{otocx_tau1_n12_p1_p0_new}(a)). In the open chain case the LMOTOC for long time is different for $\tau_1$ and $\frac{\pi}{2}-\tau_1$ (see for example $\epsilon$ and $13\epsilon$, $2\epsilon$ and $12\epsilon$, $3\epsilon$ and $11\epsilon$   in   \Cref{otocx_tau1_n12_p1_p0_new}(b)), therefore the symmetry along $\tau_1=\frac{\pi}{4}$ is absent in \Cref{finite_phase}(b). However, a symmetry along $\tau_0=\frac{\pi}{4}$ exists in both the open and the closed chain [\Cref{finite_phase}(a,b)]. We demonstrate these symmetries for the closed chain system using a toy model  of two and four spins. First, we calculate $F_x^{l,l}(n)$ for two spin system:
After the first kick ($n=1$), $F_x^{l,l}(1)=\cos(4\tau_0)$. Since the magnetization is in the direction of coupling of spins, the interaction term $H_{xx}$ (with $\tau_1$) is not involved in the state after the first kick.
After the second kick ($n=2$) LMOTOC is given by
\begin{eqnarray}
\label{L2_n2}
F_x^{l,l}(2)&=&\frac{1}{8}\Big[1-4\cos(4\tau_1) - 4\cos(4\tau_0)(-1+\cos(8\tau_1))+3 \nonumber \\
&\times&\cos(8\tau_1)+\cos(8\tau_0)(3+4\cos4\tau_1+\cos8\tau_1)\Big].
\end{eqnarray}
The symmetry along $\tau_0=\pi/4$ is evident in the above expression as $\tau_0$ and $\tau_1$ appear in the expression with a multiple of $4k$, where $k$ is an integer. Further kicking the system will also manifest the multiplicty of $4k$ with $\tau_0$ and $\tau_1$. Next, we take the toy model of four spins case.  LMOTOC after the first kick ($n=1$) will again be $F_x^{l,l}(1)=\cos(4\tau_0)$
and after the second kick($n=2$) will be 
\begin{eqnarray}
\label{L4_n2}
F_x^{l,l}(2) &=& \frac{1}{64} \Big[\cos(4\tau_0)(21-4\cos(4\tau_1)-17\cos(8\tau_1))  \nonumber \\
&+&\cos(12\tau_0)(-5 +4\cos(4\tau_1)+\cos(8\tau_1)) \nonumber \\
&+& 2(5-12\cos(4\tau_1)  + 7\cos(8\tau_1)+  \cos(8\tau_0) \nonumber \\ 
&\times& (19+12\cos(4\tau_1)+\cos(8\tau_1)))\Big]. 
\end{eqnarray}
Again we see that $F_x^{l,l}(1)$ and $F_x^{l,l}(2)$ have $\tau_0$ and $\tau_1$ are in a multiple  of $4k$, where $k$ is the integer. Therefore, LMOTOC will be same for (1) $\tau_0$ and $\frac{\pi}{2}-\tau_0$, and (2)  $\tau_1$ and $\frac{\pi}{2}-\tau_1$ and symmetric about $\tau_0=\frac{\pi}{4}$ and  $\tau_1=\frac{\pi}{4}$ in the closed chain system.
\begin{figure}[h!]
    \centering
   \includegraphics[width=\linewidth, height=.60\linewidth]{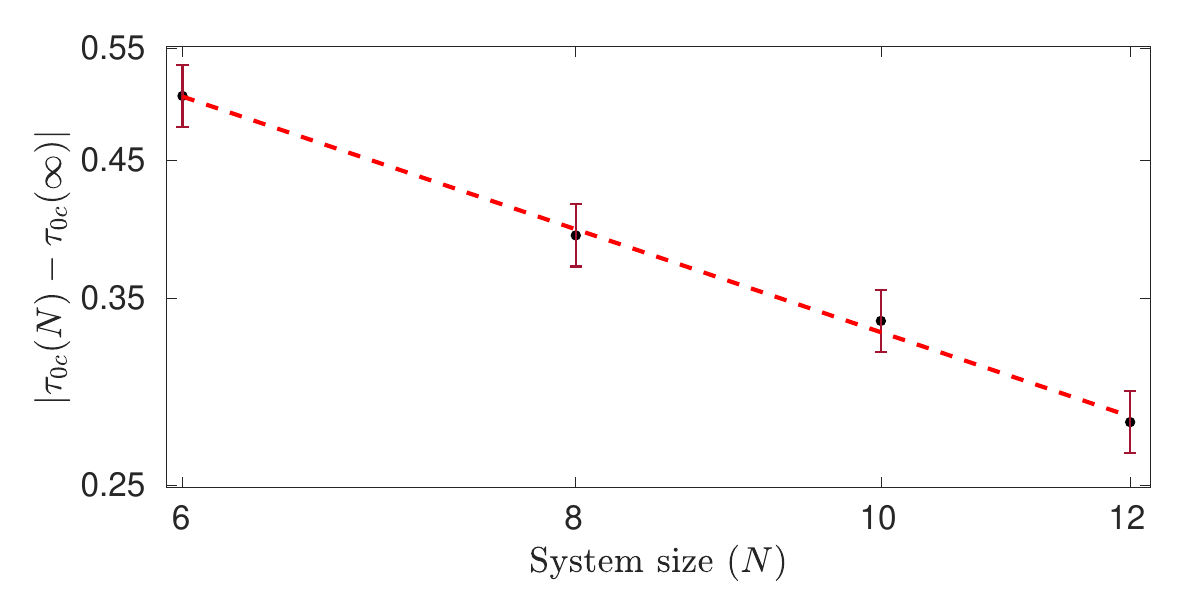}
    \caption{  Plot of the difference between finite size critical point and the infinite size critical point ($|\tau_{0c}(N)-\tau_{0c}(\infty)|$) of the phase structure of the periodic Floquet system as the function of system size (log-log plot). Black points are data points and the red dashed line is the best fit yielding the slope  $1/\nu=0.8314\pm0.1122$.}
    \label{critical_point}
\end{figure}
 In the closed chain Floquet system, the tips of the regions with $\overline{F}_x^{l,l}=0$ can be seen to be moving closer  to each other along the line $\tau_1=\frac{\pi}{4}$, with increasing the system size [\cref{finite_phase}(a)]. In the open boundary condition, the tips which start out in the upper half of the parameter space, also move downwards towards the point $(\frac{\pi}{4},\frac{\pi}{4})$ with increasing system size [\cref{finite_phase}(b)]. Behaviour of increasing the tips with increasing system size in closed chain case can be understood by the finite size effect analysis which is given as \cite{Korniss2000} 
\begin{equation}
|\tau_{0c}(N)-\tau_{0c}(\infty)|\propto N^{-1/\nu},
\end{equation}
where $\tau_{0c}(N)$($\tau_{0c}(\infty)$) is the location of the critical point on the horizontal axis of the phase structure  of the finite system size [\cref{finite_phase}(a)] (infinite system size  [\cref{four_phase}]).  $\nu$ is the transverse field exponent defined as the reciprocal of the slope of the straight line drawn from $|\tau_{0c}(N)-\tau_{0c}(\infty)|$ {\it vs} system size $(N)$ (log-log plot). As evident from  \cref{critical_point}, increasing the system size $N$ leads to closing the gap between  $\tau_{0c}(N)$ and  $\tau_{0c}(\infty)$.
In the thermodynamic limit  we expect the tips to meet at the centre, giving the diagonal lines as shown in \cref{four_phase}. Similar argument holds true for the open chain case. Hence, the time-averaged LMOTOC [$\overline{F}_x^{l,l}(T)$] for large $T$ and $N \rightarrow \infty$ can be used as an order parameter to distinguish the phases of a driven transverse field Floquet Ising model. It must be noted that the time-averaged LMOTOC does not distinguish between the ferromagnetic and the $\pi$ ferromagnetic phase or the paramagnetic and the $0\pi$ paramagnetic phase. However, these distinct phases can be identified by observing the combined eigenvalues at the edges of the phase structures of the unitary operator which is defined in \cref{U_f} and the parity operator ($P=\prod_{l}\sigma_l^z$) \cite{von2016phase,Keyserlingk2016a}. Considering the  operators $U$ and $P$ have eigenvalues $u$ and  $p$ respectively, the different phases can be distinguished by observing the eigenvalues along the outer edges of the phase diagram. The eigenvalues have protected multiplets of the form: $(u,p)$ in the paramagnetic, $[(u,p), (u,-p), (-u,p), (-u,-p)]$ in the  $0\pi$ paramagnetic, $[(u,p),  (u,-p)]$ in the ferromagnetic and   $[(u,p), (-u,-p)]$ in the $\pi$ ferromagnetic regions.
\section{Conclusion}
\label{conclusion}
We calculated the exact analytical expression for  TMOTOC  as a function of $\tau_0$ and $\tau_1$. With the help of the analytical formulation, we calculated the speed of commutator growth for the TMOTOC and compared it with those of the LMOTOC. We also analyzed the revival of the intial state and found that the TMOTOC revives back within finite time while  LMOTOC did not. Further, we study the phase structure of the traverse field Floqulationng system given by \cref{U_f} using numerical calculation of LMOTOC. We use LMOTOC defined in equation \cref{F_x} to distinguish between the paramagnetic and ferromagnetic phases of the chosen Floquet system.  Ferromagnetic and the $\pi$ ferromagnetic phase or the paramagnetic and the $0 \pi$ paramagnetic phase are distinguished by the combined eigenvalues of $U$ and parity operator $P$ along the edges of the phase structures. We numerically find the time averaged LMOTOC [$\overline{F}_x^{l,l}(T)$] for the system sizes up to $N=10$ and plot the regions of the parameter space that have $\overline{F}_x^{l,l}(T)=0$ and $\overline{F}_x^{l,l}(T)>0$ for $T=10^4$. We observe that the plot showing the critical lines of phase transition for $N \rightarrow \infty $ tends to the expected plot \cref{four_phase}. In the limit $N \rightarrow \infty $, the regions with $\overline{F}_x^{l,l}(T)>0$ for large $T$ are ferromagnetic and those with $\overline{F}_x^{l,l}(T)=0$ for large $T$ are paramagnetic.
The region in which we were discussing the phase structure, do not heat up with the large number of kicks \cite{DALESSIO201319}.
\par
OTOCs can be experimentally calculated \cite{Li2017}, and Floquet systems can be experimentally realized \cite{allen2007, Chitsazi2017}. Our study outlines the analytical calculation of the TMOTOC, its behavior with the separation between the  observables and how LMOTOC can be a useful tool to distinguish the phases of a Floquet system.

\section{Acknowledgement} 
We would like to thank V. Subrahmanyam for his fruitful suggestion in  the analytical calculation of  TMOTOC.
\appendix

\section{Calculation of transverse magnetization OTOC}
\label{Appendix1}
For the calculation of the transverse magnetization OTOC due to the local operators placed at different sites, we consider  $V=\sigma_m^z$ and $W=\sigma_l^z$. Hence TMOTOC is defined as:
 \begin{equation}
 \label{F_z_g}
F_x^{l,m}(n) =  \langle \phi_0|\sigma_l^z(n)\sigma_m^z\sigma_l^z(n)\sigma_m^z|\phi_0\rangle, 
 \end{equation}
 We transform the spin variables to fermionic creation $c_l^\dagger$ and annihilation $c_l$ operators at site $l$ by using the Jordan-Wigner transformation \cite{Jordan1928}
 \begin{equation}
S_l^x= -  \frac{1}{2}\prod_{j=1}^{l-1}(2 c_j^\dagger c_j-1)(c_l^\dagger+c_l)   \quad {\rm and}  \quad
S_l^z= c_l^\dagger c_l -  \frac{1}{2}.
\end{equation}
The operators $ c_l$ and $c_l^\dagger$ obey the the usual fermion  anticommutation rules.  The unitary operator for the closed chain is given as 
\begin{eqnarray}
U &=& \exp \Big[\frac{-i  t_1}{4} \Big( \sum_{l=1}^{N-1}(c_l^\dagger - c_l)(c_{l+1}^\dagger - c_{l+1}) \nonumber \\
    &-& (-1)^{N_F}(c_N^\dagger - c_N)(c_{l+1}^\dagger - c_{N+1})\Big) \Big] \nonumber \\
     &\times& \exp\Big[-i t_0\sum_{l=1}^{N}\big(c_l^\dagger c_l-\frac{1}{2}\big)\Big],
\end{eqnarray}
where $N_F=\sum_{l=1}^{L}c_l^\dagger c_l$ is the total number of fermions. 
We move in the momentum space using the Fourier transform of $c_l$  which is defined as
\begin{equation}
c_{q}=\frac{exp(i \frac{\pi}{4})}{\sqrt{N}}\sum_{l=1}^{N} e^{-i q l}c_l.
\end{equation}

Hence U can be written as \cite{Arul2005}
\begin{equation}
U=e^{(-i t_0 \frac{N}{2})} \prod_{q>0} \mathcal{V}_q.
\end{equation}
The operator $\mathcal{V}_q$ in the above expression has the form
\begin{eqnarray}
\mathcal{V}_q &=& \exp\Big(- i \frac{t_1}{2}[\cos(q)(c_q^\dagger c_q+c_{-q}^\dagger c_{-q})+\sin(q)(c_q c_{-q}  \nonumber \\
&+& c_{-q}^\dagger c_{q}^\dagger)] \Big)
  \exp\Big(-2 i t_0 (c_q^\dagger c_q +c_{-q}^\dagger c_{-q})\Big).
\end{eqnarray}

For $\mathcal{V}_q$, the four basis state are $|0\rangle$ , $|\pm q\rangle=c_{\pm q}^\dagger |0\rangle$,   $|-q q\rangle=c_{-q}^\dagger c_{q}^\dagger |0\rangle$

The eigenstates of $\mathcal{V}_q$ are given by
\begin{equation}
\mathcal{V}_q|\pm q\rangle=e^{\big(-\frac{t_1}{2}cos(q)- it_0 \big)}|\pm q\rangle, \nonumber
\end{equation}
\begin{equation}
\mathcal{V}_q|\pm \rangle=e^{\big(-\frac{t_1}{2} cos(q)- it_0} \big))e^{\pm i \gamma_q}|\pm \rangle,
\end{equation}
where 
\begin{equation}
|\pm \rangle = \alpha_{\pm q}|0 \rangle+ \beta_{\pm q}|-q q \rangle.
\end{equation}
 In the above equation $\alpha_{\pm q}$ and $ \beta_{\pm q}$ are given by \cref{apmq} and eq. (\ref{bpmq}), respectively.
The initial unentangled state is $|\psi_L(0)\rangle=|0\rangle^{\otimes L}$. In a Fock space, it is treated as vacuum. Time evolution operator of the fermionic annihilation operator in the momentum space is given as
\begin{equation}
\label{cqt}
c_{q}(n)=\mathcal{V}_q^{\dagger n} c_{q} \mathcal{V}_q^n=\Phi_q(n)^{*} c_q-\Psi_q(n) c_{-q}^\dagger.
\end{equation}
The expansion coefficients  $\Phi_q(n)$ and  $\Psi_q(n)$ are defined in \cref{phi} and \cref{psi}, respectively, and phase angle ($\gamma_q$) is defined in \cref{gamma}.
Let us apply the first spin operator on the initial state, we get
\begin{equation}
  S_m^z(0)|0\rangle=\Big(c_m^\dagger c_m-\frac{1}{2}\Big)|0\rangle=- \frac{1}{2}|0\rangle.
\end{equation}
In the above, $c_m^\dagger c_m$ is a number operator. The operation of the number operator on the vaccum gives zero eigenvalue. Time evolution of the spin operator at position $l$ is
\begin{eqnarray}
S_l^z(n) =  \frac{1}{N}\sum_{a,b} e^{i(a-b)l} c_a^\dagger(n) c_b(n)-\frac{1}{2},\nonumber 
 \end{eqnarray}
where $a$ and $b$ are  indices in momentum space. By using eq. (\ref{cqt}), we can write 
\begin{eqnarray}
S_l^z(n)  &=& \frac{1}{N}\sum_{a,b} e^{i(a-b)l} \Big[\Phi_a(n) c_a^{\dagger}-\Psi_a(n)^{*} c_{-a} \Big] \nonumber \\
 &\times& \Big[ \Phi_b(n)^{*} c_b- \Psi_b(n) c_{-b}^\dagger \Big] - \frac{1}{2},\nonumber \\
  &=& \frac{1}{N}\sum_{a,b} e^{i(a-b)l} \Big[ \Phi_a(n)  \Phi_b(n)^{*} c_a^{\dagger} c_b -  \Phi_a(n)  \Psi_b(n) c_a^{\dagger} c_{-b}^\dagger  \nonumber \\
  &-&  \Psi_a(n) \Phi_b(n)^{*}   c_{-a} c_b +\Psi_a(n)^{*}  \Psi_b(n) c_{-a} c_{-b}^\dagger \Big] - \frac{1}{2}. \nonumber \\
 \end{eqnarray}
Application of time evolved spin operator on the vacuum gives
 \begin{eqnarray}
 S_l^z(n)|0\rangle &=& \Big[ -\frac{1}{N}  \sum_{a,b} e^{i(a-b)l} \Phi_a(n) \Psi_b(n)  c_a^\dagger c_{-b}^\dagger \nonumber \\
 &+&  \frac{1}{N} \sum_{a}|\Psi_a(n)|^2 -\frac{1}{2}\Big]|0\rangle,
\end{eqnarray} 
and the Hermitian conjugate of the above equation is
\begin{eqnarray} \label{A12}
\langle 0 | S_l^z(n)&=&  \langle 0 |\Big[-\frac{1}{N}\sum_{p,r} e^{-i(p-r)l} \Phi_p(n)^* \Psi_r(n)^* c_{-r} c_p \nonumber \\
&+& \frac{1}{N} \sum_{p}|\Psi_p(n)|^2 -\frac{1}{2}\Big],
\end{eqnarray}
where $p$ and $r$ are indices in the momentum space. We can calculate $S_l^z(n)S_m^z(0)|0\rangle $ as 
\begin{eqnarray}
S_l^z(n)S_m^z(0)|0\rangle &=& - \frac{1}{2}\Big[ -\frac{1}{N}  \sum_{a,b} e^{i(a-b)l} \Phi_a(n) \Psi_b(n)  c_a^\dagger c_{-b}^\dagger \nonumber \\
 &+&  \frac{1}{N} \sum_{a}|\Psi_a(n)|^2 -\frac{1}{2}\Big]|0\rangle.
\end{eqnarray}
Applying the third spin operator $S_m^z(0)$ on the state 
\begin{widetext}
$S_l^z(n)S_m^z(0)|0\rangle$  we get
\begin{eqnarray}\label{A14}
S_m^z(0)S_l^z(n)S_m^z(0)|0\rangle   &=& -\frac{1}{2} \Big[\frac{1}{N}\sum_{x,y} e^{i(x-y)m}c_x^\dagger c_y - \frac{1}{2} \Big] 
 \Big[ -\frac{1}{N}  \sum_{a,b} e^{i(a-b)l} \Phi_a(n) \Psi_b(n)  c_a^\dagger c_{-b}^\dagger 
 +  \frac{1}{N} \sum_{a}|\Psi_a(n)|^2 -\frac{1}{2}\Big]|0\rangle,\nonumber \\
&=& -\frac{1}{2} \Big[-\frac{1}{N^2}\sum_{x,y,a,b} e^{i(x-y)l}  e^{i(a-b)l}
  \Phi_a(n) \Psi_b(n)  \Big(c_x^\dagger c_{-b}^\dagger \delta(a,y)  
-c_x^\dagger c_{a}^\dagger \delta (-b,y)\Big) \nonumber \\ 
&+&  \frac{1}{2N}\sum_{a,b} e^{i(a-b)l} \Phi_a(n) \Psi_b(n)  c_a^\dagger c_{-b}^\dagger 
- \frac{1}{2} \Big(\frac{1}{N} \sum_{a}|\Psi_a(n)|^2 - \frac{1}{2} \Big)\Big]  |0\rangle,
\end{eqnarray}
\end{widetext}
where $x$ and $y$ are the indices in the momentum space. Now we take the scalar product of the states given by eq. (\ref{A12}) and eq. (\ref{A14}) and get TMOTOC as
\begin{widetext}
\begin{eqnarray}
F_x^{l,m}(n) &=& 2^4 \langle 0|S_l^z(n) S_m^z(0)S_l^z(n)S_m^z(0)|0\rangle, \nonumber \\
&=&-2^3\langle 0| \Big[ -\frac{1}{N}\sum_{p,r} e^{-i(p-r)l} \Phi_p(n)^{*} \Psi_r(n)^* c_{-r} c_p 
+ \frac{1}{N} \sum_{p}|\Psi_p(n)|^2 -\frac{1}{2} \Big]
  \Big[-\frac{1}{N^2}\sum_{x,y,a,b} e^{i(x-y)m}  e^{i(a-b)l}\Phi_a(n) \nonumber \\
 &\times& \Psi_b(n) \Big(c_x^\dagger c_{-b}^\dagger \delta(a,y)  -c_x^\dagger c_{a}^\dagger \delta (-b,y)\Big)  
+ \frac{1}{2N}\sum_{a,b} e^{i(a-b)l} \Phi_a(n) \Psi_b(n)  c_a^\dagger c_{-b}^\dagger 
-\frac{1}{2} \Big( \frac{1}{N} \sum_{a}|\Psi_a(n)|^2 -\frac{1}{2} \Big) \Big]  |0\rangle, \nonumber \\
&=& -2^3 \Big[\frac{1}{ N^3} \Big( \sum_{p,a,r,x,y}e^{-i(p-a)l}e^{i(x-y)m}|\Psi_r(n)|^2 \Phi_p^*(n) \Phi_a(n) \delta(p,x) \delta(a,y)
- e^{i(r+a)l}e^{i(x-y)m}  \Psi_r(n)^* \Phi_p^*(n) \Phi_a(n)  \nonumber \\
&\times& \Psi_{-p}(n) \delta(-r,x) \delta(a,y)- e^{-i(p+b)l}e^{i(x-y)m} \Psi_{b}(n)  \Psi_{-a}(n)^*  \Phi_{p}(n)^* \Phi_{a}(n)\delta(p,x) \delta(-b,y)+ e^{i(r-b)l}e^{i(x-y)m}\Psi_{b}(n)    \nonumber  \\
 &\times& \Psi_r(n)^* |\Phi_a(n)|^2 \delta(p,a)\delta(-b,y)\Big) 
 - \frac{1}{ 2N^2}  \sum_{p,r} \Big(|\Psi_p(n)|^2 |\Phi_r(n)|^2- \Psi_{-p}(n)  \Psi_r(n)^* \Phi_{p}(n)^* \Phi_{-r}(n) \Big) \nonumber \\
 &-& \frac{1}{2} \Big(\frac{1}{N}  \sum_{p}| \Psi_p(n)|^2-\frac{1}{2}\Big)\Big(\frac{1}{N}  \sum_{a}| \Psi_a(n)|^2-\frac{1}{2}\Big) \Big].
\end{eqnarray}
Since, the term 
\begin{eqnarray}
  \frac{1}{ 2N^2}  \sum_{p,r} \Big(|\Psi_p(n)|^2 |\Phi_r(n)|^2- \Psi_{-p}(n)  \Psi_r(n)^* \Phi_{p}(n)^* \Phi_{-r}(n) \Big)
 +\frac{1}{2} \Big(\frac{1}{N}  \sum_{p}| \Psi_p(n)|^2-\frac{1}{2}\Big)\Big(\frac{1}{N}  \sum_{a}| \Psi_a(n)|^2-\frac{1}{2}\Big) \nonumber 
 \end{eqnarray}
is constant for all number of kicks $(n)$ and system size $(N)$ which comes out to be $\frac{1}{2^3}$. Since $a$ and $b$ are  dummy indices, we replace it by $q$.
Hence, the final formula of TMOTOC is 
\begin{eqnarray}
 F_z^{l,m}(n) &=& 1- \Big(\frac{2}{ N}\Big)^3 \sum_{p,q,r} \Big( e^{i(p-q)(m-l)}|\Psi_r(n)|^2 \Phi_p^{*}(n) \Phi_q(n)- e^{i(-r-q)(m-l)}  \Psi_r(n)^{*} \Phi_p^{*}(n)  \Phi_q(n) \Psi_{-p}(n) \nonumber \\
&-& e^{i(p+q)(m-l)} \Psi_{q}(n)  \Psi_{r}(n)^{*}  \Phi_{p}(n)^{*} \Phi_{-r}(n)+ e^{i(q-r)(m-l)}   
  \Psi_{q}(n)\Psi_r(n)^* |\Phi_p(n)|^2 \Big).  
\end{eqnarray}
Now, we take a special case in which both the operators are the same local operator i.e.  $W=\sigma_l^z$ and $V=\sigma_l^z$. Then the formula becomes
\begin{eqnarray}
 F_z^{l,m}(n)&=& 1- \Big(\frac{2}{ N}\Big)^3 \sum_{p,q,r}  \Big( |\Psi_r(n)|^2 \Phi_p^*(n) \Phi_q(n)  
-\Psi_{-p}(n) \Psi_r(n)^* \Phi_p^*(n) \Phi_q(n)\nonumber  \\
 &-&  \Psi_q(n)  \Psi_r(n)^*  \Phi_{p}(n)^* \Phi_{-r}(n)+ \Psi_{q}(n)  \Psi_r(n)^* |\Phi_p(n)|^2 \Big).  
\end{eqnarray}
\end{widetext}


\section{Calculation of LMOTOC}
\label{Appendix2}
 Let us attempt to find the analytical expression of LMOTOC so that we calculate the phase structure of LMOTOC for higher system size. After moving some steps in analytical calculation of LMOTOC, we realize that the analytical expression of LMOTOC will take longer time than the numerical calculation. A few initial steps of our calculation of LMOTOCs are given below:
$S_l^x$ in the form of raising and lowering operator:
\begin{eqnarray}
S_l^x&=&\frac{1}{2}[S_l^{+}+S_l^{-}] \nonumber \\
&=& \frac{1}{2} \exp\Big [-\pi i \sum_{j=1}^{l-1}c_j^\dagger c_j\Big](c_l^\dagger+c_l)  \nonumber \\   
&=& - \frac{1}{2} \prod_{j=1}^{l-1}(2 c_j^\dagger c_j-1)(c_l^\dagger+c_l)  \nonumber  
\end{eqnarray}
The above equation is written by using the relation 
\begin{eqnarray}
S_l^{+}=c_l^\dagger \exp[\pi i \sum_{j=1}^{l-1}c_j^\dagger c_j], \nonumber \\ 
S_l^{-}=  \exp[-\pi i \sum_{j=1}^{l-1}c_j^\dagger c_j]c_l.  \nonumber
\end{eqnarray}
Now, we move in the momentum space by doing the Fourier transform of $c_l$ and $c_l^\dagger$. Hence $S_l^x$ in momentum space can be written as: 
\begin{eqnarray}
S_l^x &=&-\frac{1}{2} \prod_{j=1}^{l-1}\Big[2 \sum_{q_{j},p_{j}} \frac{1}{L}\exp[i(p_j-q_j)j] c_{q_j}^\dagger c_{p_j}-1\Big] \Big[\sum_{r}\Big(\frac{1}{\sqrt{L}} \nonumber \\ 
&\times& \exp(\frac{-i \pi}{4})\exp(irl) c_r^\dagger + \frac{1}{\sqrt{L}} \exp(\frac{i \pi}{4})\exp(-irl) c_r\Big)\Big] \nonumber \\  
\end{eqnarray}
For the calculation of  the time evolution of $S_l^x$ {\it i.e.}, $S_l^x(n)$ we have to compute the time evolution of all the operators in the string of length $L$. Time evolution of such a large term will be too much complicated and unfruitful for our purpose because the calculation of LMOTOC involving product of four operators will be too much to handle.


\bibliographystyle{apsrev4-1}
\bibliography{Floquet_phase}
\end{document}